\documentclass[manuscript,screen]{acmart}
\AtBeginDocument{%
  \providecommand\BibTeX{{%
    \normalfont B\kern-0.5em{\scshape i\kern-0.25em b}\kern-0.8em\TeX}}}

\setcopyright{rightsretained}

\acmConference[Cross-Sensory Futures at CHI '26]{April 13, 2026}{Barcelona, Spain}
\begin{document}

\title{New Cross-Sensory Approach to Designing Restorative Virtual Environments}

\author{Rachel Masters}
\email{ramast1@colostate.edu}
\orcid{0000-0002-3857-6537}
\affiliation{%
  \institution{Colorado State University}
  \city{Fort Collins}
  \state{Colorado}
  \country{USA}
}

\author{Francisco Ortega}
\email{f.ortega@colostate.edu}
\orcid{0000-0002-2449-3802}
\affiliation{%
  \institution{Colorado State University}
  \city{Fort Collins}
  \state{Colorado}
  \country{USA}
}

\renewcommand{\shortauthors}{Masters and Ortega}

\begin{abstract}
    Virtual reality (VR) nature immersion is an increasingly popular field of research due to its potential to help people who do not have access to real nature. There are many questions surrounding how virtual forests can be designed to effectively reduce stress and restore attention. Many of these questions relate solely to visual aspects, but more recent literature has started exploring multisensory experiences. In these experiences, senses are treated as additive; however, certain results from the current literature may indicate that there are more complex, cross-sensory interactions occurring. For example, adding sound to visuals can increase stress reduction potential, but certain natural sounds can feel threatening if they are out of place within the virtual nature scene. Overall, cross-sensory interactions in VR nature environments (VNEs) are underexplored and challenge our current understanding of multisensory VNEs, and future explorations of these interactions are essential for designing optimal VNEs for stress reduction.
\end{abstract}



\maketitle

\section{Introduction}
Forest bathing is a term used to describe the practice of nature immersion for stress reduction and mental resource restoration. Stress Recovery Theory describes how nonthreatening natural environments reduce stress by decreasing negative affect and increasing positive affect ~\cite{ulrich1983aesthetic}. Attention Restoration Theory describes how nature restores depleted cognitive resources through elements that naturally maintain our interest ~\cite{kaplan1989experience}. Researchers are actively working to understand what qualities of nature make it restorative, both for nature conservation purposes ~\cite{SPANGENBERGER2024104964} and for the design of restorative parks ~\cite{liang2024urban} that can help people in high stress environments like cities and hospitals ~\cite{zandi2025health}.

Virtual reality (VR) is becoming a very useful tool for both of these cases. VR is excellent for studying the design of real world parks because it can simulate granular differences in nature environments that would be too cumbersome to control in a real nature environment~\cite{liang2024urban}, such as testing certain sounds or controlling the number of different plants. VR also has potential as a virtual nature supplement when access to real nature is lacking, promoting stress reduction in the types of high stress environments that have the least nature access, including nursing homes ~\cite{gruber2022vr} and hospitals ~\cite{birrenbach2022virtual} among other confined environments.

While VR presents great opportunities for controlled nature simulations, it also presents many unknowns surrounding the optimal design of VR nature environments (VNEs) for stress reduction and mental resource restoration. The primary focus of current literature on VNE design is on visual aspects of the environment ~\cite{masters2022virtual,marselle2021pathways,li2023eye,olvera2021associations,nicoly2024restorative}; however, multisensory VNE design is of increasing interest to researchers for multiple reasons. The primary reason is that real nature is multisensory, so incorporation of multiple senses into VNEs increases presence which can increase stress reduction potential ~\cite{jiang2016effect}. A secondary reason is that cross-sensory interactions may affect stress reduction in ways not yet understood. For example, Prospect-Refuge Theory indicates that aesthetic preference is linked to survival instinct, and therefore linked to a preference for environments that provide shelter (refuge), while also offering a clear view of the surrounding area (prospect) ~\cite{appleton1975experience}. While the theory focuses on visuals, it is unclear if the theory could extend to certain sounds such as rustling, which could just be wind in a non-threatening environment, or an animal in a threatening environment. Overall, more work is needed on multisensory effects, and this work needs to consider cross-sensory interactions.

\section{Current Multisensory Research}
Research on multisensory VNEs is growing but limited. For example, while audio has been the most studied addition to nature visuals, even audio has design complexities that affect perception of visuals, and the topic is an active area of research~\cite{aletta2021human}. The other senses are even less explored but are becoming more prevalent, demanding more attention as technology improves ~\cite{masters2023multisensory}. One systematic review indicates that the inclusion of smell improves experience, and there is more research on smell as scent technologies are improving ~\cite{lopes2024audio}. Other new studies on using scent in VNEs have also found scent to be a positive addition to audio-visual VNEs ~\cite{lopes2024stop,ascone2025multi}, but have not explicitly explored cross-sensory interactions. An article by Berdejo-Espinola et al. mentions that "feeling the comforts and discomforts of the wind, temperature and scents associated with nature — changed VR users’ perceptions of nature" ~\cite{berdejo2024virtual}, indicating an importance of combined senses for perception that needs more exploration. Song et al. did an exploration into comparing environments with different combinations of the five senses, which provided some insight into what sensory combinations are more effective than others, but it mainly focused on additive effects of different senses instead of cross-sensory interactions ~\cite{song2024effects}. Another systematic review on the current literature indicates the need for multisensory research frameworks that approach cross-sensory interactions, as the current literature mainly observes isolated senses~\cite{du2025towards}. Most importantly, the exploration of cross-sensory interactions have implications for the applications of VNEs, as recent research has shown that multisensory VNEs have the potential to help PTSD patients ~\cite{de2023using,de2025instrumenting,lopes2025subjective}.

\section{Position: Call for Cross-Sensory Approach to Multisensory VNE Research}
The current literature on multisensory VNEs indicates that audio, visual, scent, and touch components of an environment all have a role in stress reduction and restoration. While research has indicated that each sense can have added positive effects when designed correctly, the interaction between the senses is much more complex and less understood. To engineer optimally restorative VNEs, future research avenues need to explore how cross-sensory interactions can be used both to increase immersion and enrich user experience.

\section{Future Avenues of Cross-Sensory VNE Research}
Every area of multisensory VNE research can benefit from the examination of cross-sensory interactions.

\subsection{Audio and Visual Interactions}
Audio-visual environments are currently the most researched type of multisensory VNE; however, few papers have explicitly explored how the design of audio affects visual perception of the environment, or how the design of visuals affect perception of audio. A major interaction that needs studied more is the interaction between crafted soundscapes and intentional visuals. This can be done by selectively combining specific audio and visual components that match or do not match, testing user perception of a change in one component at a time. Conducting this research will provide further insights into what sounds or visuals can seem intimidating when not appropriately matched to an environment, such as how certain kinds of rustling may sound like animal presence when an animal is not in the environment, leading to anticipation or fear of an unseen threat.

\subsection{Audio-Visual and Smell Interactions}
Due to limited technology, smell interactions have been tested less; however, a huge part of nature's benefits come from the inhaled microbes that are in soil and plants ~\cite{haahtela2013biodiversity}. Therefore, there is ample opportunity to test how smell affects perception of audio-visual factors, and while scent technology is still improving ~\cite{lopes2024audio}, there are other ways to test scents. This can be split into scents of things in nature, like burning a candle with eucalyptus or lavender oil, and scents of present natural items, like bringing samples of soil, pine needles and cones, or dried leaves into the lab environment. The multisensory testing strategy would be to compare smell and no smell to see if there is a difference in stress reduction. The cross-sensory testing strategy would be to investigate people's expectation of scent in the environment, and compare scents that do and do not fit the audio-visual environment to see how perception is affected. Conducting this research will provide insights into the importance of scent interactions for an immersive VR experience, and the results of this research could provide insights into the future design of scent technologies.

\subsection{Audio-Visual and Touch Interactions}
While it is challenging to touch nature in VR, two interesting potential cross-sensory interactions to explore are wind and temperature. These can be controlled using fans and changing the room temperature, and the testing strategy is the same as testing scents. This avenue of research is especially interesting because in past studies conducted by our research group, participants sat near an air vent, and a few mentioned the breeze enhancing the visuals of the nature environment ~\cite{masters2024impact}. Given that nearly all of the VR nature assets have some sort of wind implemented, the effects of environments with and without wind of varying speeds and temperatures on perception is underexplored.

\section{Conclusion}
Multisensory VNE design is a new and emerging research area with many questions and ample potential. Cross-sensory interaction opens even more questions that need investigated and challenges our current understanding of the roles of the different senses for stress reduction and attention restoration. It is the combination of senses that can make a VNE stress reducing or stress inducing, and it is the combination of senses that makes nature more complex and interesting, potentially affecting the attention restoration capacity of an environment.

\section{Acknowledgements}
This material is based upon work supported by the National Science Foundation Graduate Research Fellowship Program under Grant No. 23605. Any opinions, findings, and conclusions or recommendations expressed in this material are those of the author(s) and do not necessarily reflect the views of the National Science Foundation

\bibliographystyle{ACM-Reference-Format}
\bibliography{sample-base}

\appendix

\end{document}